\documentclass[a4paper,twocolumn,amsmath,amssymb,aps,prl,10pt,superscriptaddress]{revtex4-1}

\usepackage{amsmath}
\usepackage{amssymb}
\usepackage{graphicx}
\usepackage{color}

\pdfoutput=1

\begin{document}
\title{Majorana boundary modes in Josephson junctions arrays with gapless bulk excitations}
\author{M.\ Pino}
\affiliation{Department of Physics and Astronomy, Rutgers The State University of New Jersey, 136 Frelinghuysen rd, Piscataway, 08854 New Jersey, USA}
\author{ A.\ M.\ Tsvelik}
\affiliation{Department of Condensed Matter Physics and Materials Science, Brookhaven National Laboratory, Upton, New York 11973-5000, USA}
\author{L.\ B.\ Ioffe}
\affiliation{Department of Physics and Astronomy, Rutgers The State University of New Jersey, 136 Frelinghuysen rd, Piscataway, 08854 New Jersey, USA}
\affiliation{CNRS, UMR 7589, LPTHE, F-75005, Paris, France.}
\begin{abstract}
The search for Majorana bound states in solid-state physics has been limited to materials which display a gap in their bulk spectrum. 
We show that  such unpaired states appear in certain quasi-one-dimensional  Josephson junctions arrays with gapless bulk
excitations. The bulk modes mediate a coupling between Majorana bound states via the  Ruderman-Kittel-Yosida-Kasuya mechanism.
As a consequence, the lowest energy doublet acquires a finite energy difference. For realistic set of parameters this energy splitting   remains much smaller than the energy of the bulk eigenstates even for short chains of length $L \sim 10$.
\end{abstract}
\maketitle

\paragraph{Introduction.}

An intensive search for Majorana fermions ~\cite{Ma37} is underway in solid-state devices~\cite{Al12}.
The vast majority of the proposals consist in zero energy boundary modes in materials with a gaped bulk spectrum.
A spin-less superconducting wire or topological insulator in two or three dimensions fall in this category~\cite{Ki01, Mo12,Ha10}.

We propose an alternative approach for the observation of Majorana fermions in Josephson Junctions Arrays (JJA). 
We will show that certain quasi-one-dimensional JJA can display Majorana zero modes at their boundaries. 
These modes are protected from mixing with higher energy excitations although bulk spectrum is not gapped. 
The existence of low energy Majorana could then be proved by spectroscopy\ \cite{BeIo12,BePa14}.

In this letter, we first explain the JJA system and how to model it with
an Ising-like Hamiltonian. Then, a qualitative argument is employed to obtain the low-energy effective theory using Majorana boundary modes.
Numerical results will confirm the validity of this effective theory and show that Majorana modes are indeed protected.
Finally, we discuss problems that may arise in the experimental realization of our proposal.

\begin{figure}[!t]
\begin{center}
 \includegraphics[width=8.0cm]{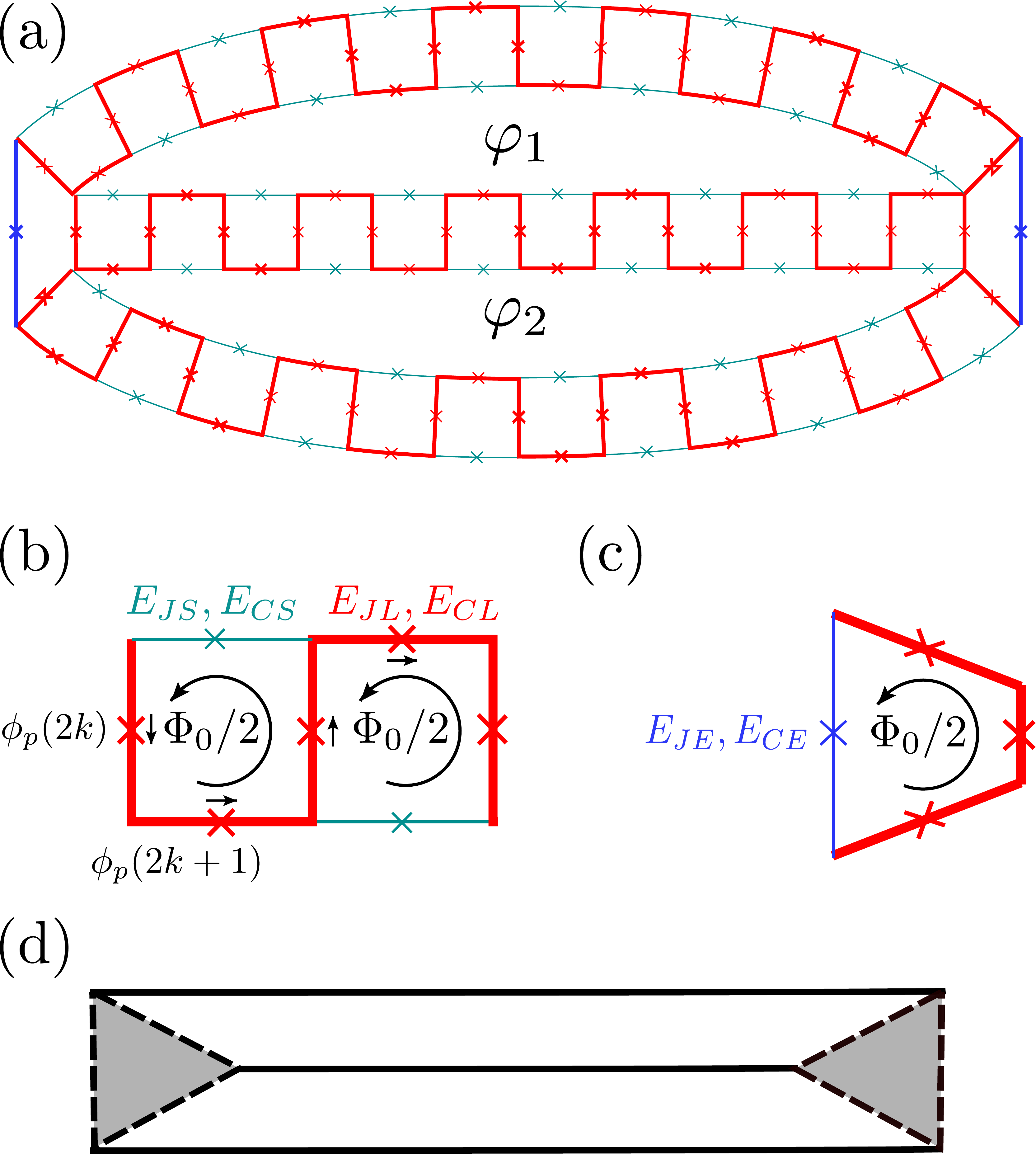}     
\end{center}
\caption{(a) Three ladders of Josephson junctions coupled together. The two large loops are threaded by magnetic fluxes $\varphi_1$ and $\varphi_2$. 
(b) Unit cell of one ladder. Josephson energy of ''large'' junction (red) is larger than the one of a ''small'' junction (green), $E_{JL}>E_{JS}$.
All the cells are threaded by a magnetic flux equal to half of the flux quantum. 
(c) Loop at the left boundary (the one at right is symmetric). The Josephson coupling for the blue 
junction at the chains end is denoted as $E_{JE}$.
(d) For small temperature and near a phase transition, the three Josephson ladders system maps to three Ising chains with transverse field (solid lines)
and coupling between their ends (dotted lines). Majorana zero modes are located at the chains boundaries (dark regions).  
}\label{Fig1}
\end{figure}

\paragraph{Experimental set-up}
We consider three identical ladders of Josephson junctions coupled together as in Fig.\ (\ref{Fig1}a)\ \cite{BeSa12}.
Each ladder has a unit cell with ''large'' and  ''small'' junctions, Fig.\ (\ref{Fig1}b). Their corresponding Josephson energies are $E_{JL}$ and $E_{JS}$, where $E_{JL}>E_{JS}$. 
The three ladders are closed at the ends by a junction with Josephson energy $E_{JE}$.
We assume that charging energies are much smaller than Josephson ones for all the junctions.
All the closed circuit in the ladders are at full frustration, that is, they are threaded by a magnetic flux equal to half of the flux quantum.
The two larger loops are threaded by magnetic fluxes $\varphi_1$ and $\varphi_2$, respectively.
For ladder $p$, the phase difference in the left vertical and horizontal red junctions in the loop $k$ are denoted by $\phi_p(2k)$ and  $\phi_p(2k+1)$, respectively (see Fig.\ \ref{Fig1}b) .

At full frustration, the properties of one infinite chain are invariant under translation by one small closed loop and reflection through horizontal axes. 
We analyze the energy of one unit cell as a function of the phase difference in the ''large'' junctions:
\begin{align}
 \mathcal{U}_{p,k}&=-E_{JL}\ \Big\{ \cos\left[\phi_p(2k)\right]+\cos\left[\phi_p(2k+1)\right]\Big\}\nonumber\\
                     +&E_{JS}\ \cos\left[\phi_p(2k)+\phi_p(2k+1)+\phi_p(2k+2)\right],\label{eq:JE}
\end{align}
where $ \mathcal{U}_{p,k}$ is the energy of the loop $k$ of ladder $p$.
The total energy has a global $\mathbb{Z}_2$ symmetry given by $\phi_p(i)=-\phi_p(i)$ in each junction $i$. In the regime $E_{JL}\gg E_{JS}$, the ground-state corresponds to 
all phases equal to zero $\phi_p(i)=0$ and the symmetry is preserved. However, a broken symmetry phase occurs for $E_{JL}\ll E_{JS}$, as in this regime
the ground-state acquires a $\phi_p(i)\ne 0$. The critical point is located near $E_{JL}/ E_{JS}\sim 5$\ \cite{BeSa12}. 

The specific details of the system are irrelevant 
near the phase transition and the properties of one ladder are described by an Ising chain with transverse field. 
The mapping can be written explicitly by taking each ''large junction'' as a 1/2-spin with component $\sqrt{1-\phi^2_p(i)}$ in the direction of the field.
Near the phase transition, only the lowest non-zero order in the phase differences in ''large'' junctions are relevant. 
Then, the first term in Eq.\ (\ref{eq:JE}) represents the field contribution and the second the Ising coupling between nearby spins.

The junctions located at the ends of the chains, blue crosses in Fig.\ (\ref{Fig1}d), couple the three ladders: 
\begin{align}
\mathcal{U}_{\rm c}=E_{JE}\ &\Big\{ \cos\left[\phi_1(1)+\phi_2(1)+\phi_3(1)\right]\nonumber\\
         &+\cos\left[\phi_1(L)+\phi_2(L)+\phi_3(L)\right]\Big\},\label{eq:JEc}
\end{align}
where $L$ is the double of the number of squares in each ladder. This contribution gives an extra coupling between spins at the boundaries in the Ising model.

\paragraph{Effective model.}
Near the phase transition, the Hamiltonian of the three ladders of JJA maps to three Ising chains with transverse magnetic field and coupling between their ends $H=\sum_p H_{p}+H_{\rm c} $, where:
\begin{align}
 H_{ p}=&-J\sum_{k=1}^{L-1} \sigma_{p}^{x}(k)\sigma_{p}^{x}(k+1)-h\sum_{k=1}^L \sigma_{p}^{z}(k),\label{eq:3Isinga}\\
 H_{\rm c}=&-J_{c}\sum_{p=1}^3 \sigma_{p}^{x}(1)\sigma_{p+1}^{x}(1) + \sigma_{p}^{x}(L)\sigma_{p+1}^{x}(L).\label{eq:3Isingb}
\end{align}
The index $p=1,2,3$ is the chain number and it has periodic boundary conditions. The
length of each chain is $L$ and thus, the total number of sites of the three chains is $N=3L$.
The operators $\sigma_{a}^{z}(i),\sigma_{a}^{x}(i),\sigma_{a}^{z}(i)$ are Pauli matrices acting on the Hilbert space of the spin at site $i$ of chain $p$.
The parameters $J$ and $h$ control the usual Ising interaction and magnetic field in the transverse axis, respectively. 
They can be related to the original Josephson coupling energies as $J/h\sim E_{JL}/E_{JS}$
The value of $J_{\rm c}$ sets the strength of the coupling between different chains.
It can be controlled by the junctions located at ends of the Josephson ladders $J_c/J\sim E_{JE}/E_{JL}$.  
We are interested in the regime $h/J\sim 1$ and $J_c/J\sim 1$.
A schematic representation of the Hamiltonian appears in Fig.\ (\ref{Fig1}d). 

In Eqs. (\ref{eq:JE},\ref{eq:JEc}), we have implicitly assumed that quantum fluctuations in the phase of the ''small'' junctions are zero.
In practice, these fluctuations are small but non-zero. We have also neglected any noise in the flux threading superconductor loops.
These two contribution represent sources of incoherent noise for our effective Hamiltonian. We will comment later on how those contributions affect our results.

We express the low-energy degrees of freedom for Hamiltonian Eq.~(\ref{eq:3Isinga},\ref{eq:3Isingb}) in terms of Majorana fermions. We use a multichannel version of the Jordan-Wigner transformation~\cite{Cr13}. 
This mapping requires to enlarge the Hilbert space with an extra 1/2-spin. 
Operators acting on this spin are denoted by $\sigma^x(0)$, $\sigma^y(0)$ and $\sigma^z(0)$.
The original spin operators are mapped to fermions via:
\begin{align}\label{eq:fermions}
c_p^\dagger(k)=\eta_p (-1)^{\sum_{j<k}n_p\left(j\right)}\sigma_{p}^+\left(k\right),
\end{align}
where $n_p(j)=\left[\sigma_j^z(p)+1\right]/2$. The $\eta_p$ acts on the added 1/2 fermion. 
They are: 
\begin{align}
\eta_1  &= \sigma^x(0)(-1)^{N_2+N_3}, \nonumber\\
\eta_2  &= \sigma^y(0)~(-1)^{N_1+N_3}, \nonumber\\
\eta_3  &= \sigma^z(0)~(-1)^{N_1+N_2}, \nonumber
\end{align}
where $N_p$ is the total number of fermions in chain $p$. The operators defined in Eq~(\ref{eq:fermions}) fulfill the fermionic algebra. 
Finally, Majorana fermions are defined as $\psi_p(2k-1)=c_p(k)+c^\dagger_p(k)$ and $\psi_p(2k)=i\left[c_p(k)-c^\dagger_p(k)\right]$.

We focus on the regime in which each of the Ising chains are critical, $J=h$. In this case,
our model can be written as:
\begin{align}
 &iH=J\sum_{j=1}^{2L} \vec{\psi}(j)\vec{\psi}(j+1)+\label{Eq.c}\\
&J_c\sum_{abc} \epsilon_{abc}\eta_{b}\eta_{c} \left[\psi_b(1)\psi_c(1)+\psi_b(2L)\psi_c(2L)(-1)^{N_b+N_c}\right],\nonumber
\end{align}
where $\vec{\psi}(j)=(\psi_1(j),\psi_2(j),\psi_3(j))$.
The coupling at the left boundary is $\vec{S}\cdot(\vec{\psi}(1)\times \vec{\psi}(1))$, where $S^a=(i/2)\epsilon_{abc}\eta_a\eta_b$. 
If only this coupling is considered, the Hamiltonian describes an over-screened two channel Kondo (2CK)  model ~\cite{Ts13,Co95,Al14,Gi13}. The role of the impurity is played by
the 1/2-spin $\vec{S}$ and the relevant degrees of freedom of the conduction electrons are described by the bulk spins of the three Ising chains. 
At temperatures below the Kondo temperature $T_K \sim J\exp(-\mbox{const} J/J_c)$  the two-channel Kondo model displays the universal quantum critical behavior. At the Quantum Critical point (QCP) of the 2CK model possesses a finite ground state entropy $\ln\sqrt 2$ which originates from the presence of a zero energy Majorana mode. 
This mode presents what is left from the impurity spin S=1/2. The leading irrelevant operator at the QCP describes a coupling of this mode to the product of three bulk Majorana at the impurity point. 
 
It stands to reason that at $L \rightarrow \infty$ model (\ref{Eq.c}) describes two independent 2CK models. Formally such decoupling is achieved by declaring $(-1)^{N_a}\eta_a$ as an independent Majorana fermion. Then its anticommutator with $\eta_a$ is zero on average. At finite $L$  Hamiltonian (\ref{Eq.c}) possesses an additional energy scale generated by a tunneling between the Majorana modes located at different ends of the system. 
Below this scale one has to see deviations from the two-channel Kondo physics.
Although model (\ref{Eq.c}) is not integrable, we can analyze qualitatively the low energy theory starting from the strong coupling limit  $J_c/J =0$.
There are  Majorana  fermions at each end of the chains, $\mu_L=\psi_1(1)\psi_2(1)\psi_3(1)$ and $\mu_R=\psi_1(2L)\psi_2(2L)\psi_3(2L)$, which commute with the strong coupling Hamiltonian.
That implies a two-fold degeneracy of the whole spectrum. Indeed, the occupancy of the fermionic state $f=[\mu_L-i\mu_R]$ does not affect the energy.
We notice that the fermion $f$ is constructed from Majorana at different ends of the chain.

If one adopts the hypothesis of independence of two 2CK QCPs at $L \rightarrow \infty$ and consider  the finite size effects as a perturbation their contribution to the low energy dynamics can be extracted from the effective Hamiltonian computed as~\cite{Co95}:
\begin{align}
H_r\sim T_K^{-1/2} &[\mu_L\psi_1(2)\psi_2(2)\psi_3(2)+ \nonumber\\
 &\mu_R\psi_1(2L-1)\psi_2(2L-1)\psi_3(2L-1)]\label{eff}.
\end{align}
We expect that Majorana at each end couple via RKKY mechanism through the low energy excitations~\cite{Ru54,Ka56,Yo57,Er15}. Indeed, integrating over the fermions in (\ref{eff}) we obtain the effective Hamiltonian for the Majorana modes:  $H_{eff}=t~\mu_L\mu_R$ where $T \sim T_K^{-1}\int G(L,t)^3dt\sim T_K^{-1} (v_F/L)^{2}$. The Green function 
of the bulk Majorana is 
\begin{align}
G(x,t) = &e^{ik_F x}\frac{v_F\pi}{2L\sin\Big[\pi( x + i t v_F)/2L\Big]} \nonumber \\
        +& e^{-ik_F x}\frac{v_F\pi}{2L\sin\Big[\pi( x - i t v_F)/2L\Big]},
\end{align}
where $k_F = \pi/2$ is the Fermi wave vector  and $v_F$ is the Fermi velocity ~\cite{Co95}. In this simpler case, 
the degeneracy of the spectrum is lifted by a factor $T\sim L^{-2}$. The effective Hamiltonian can be written as 
\begin{align}
 H_{eff}=T~(f^\dagger f-\frac{1}{2}).
\end{align}

If this result holds for the full model Eq.~(\ref{eq:3Isinga},\ref{eq:3Isingb}), the existence of Majorana fermions in the JJA discussed above would be signaled by 
a mode with energy going as $L^{-2}$.
The bulk eigenstates correspond to that of the Ising chain with transverse field at the critical point~\cite{Pf70}, 
whose energy levels scale as $\epsilon\gtrsim  1/L$.
Then, for sufficiently long chains the unpaired Majorana are protected. 
We are going to check that this is actually the case  for the full model  Eqs.~(\ref{eq:3Isinga},\ref{eq:3Isingb}) with parameters $J=h=J_c$. 
In fact, it turns out that for the parameters chosen the low energy doublet is still protected even in the case of short chains.

\paragraph{Numerical results.}

\begin{figure}[t]
\begin{center}
 \includegraphics[width=8.2cm]{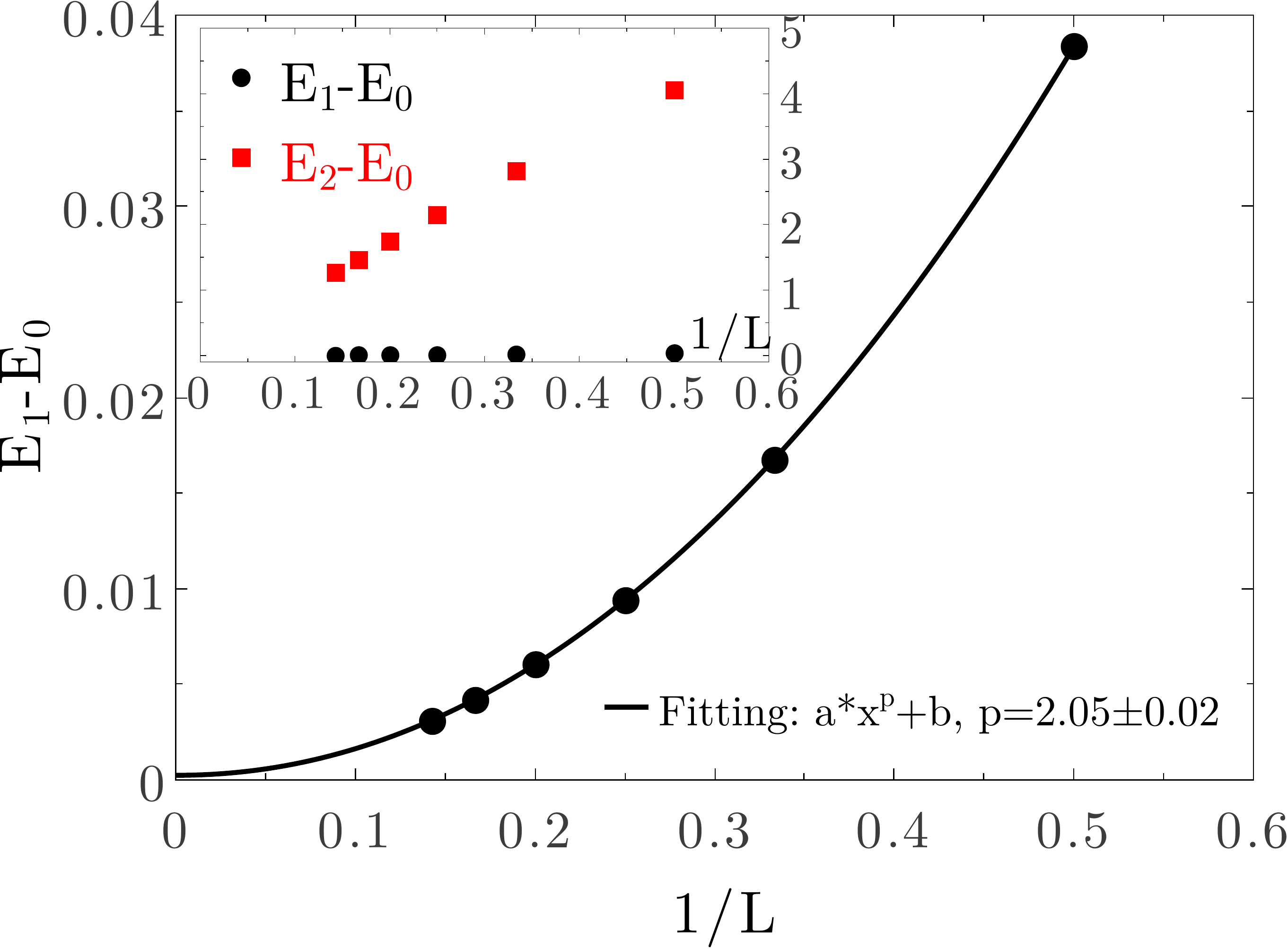}     
\end{center}
\caption{Difference between the energy of the first excited and ground states, $E_1-E_0$, of the three Ising chains Eqs.\ (\ref{eq:3Isinga},\ref{eq:3Isingb})  as a function of the inverse of 
the chains length $1/L$. Each chain is at the critical point $h=J$. The strength of the coupling between chains is $J_{c}=J$.
The size of the system goes from $N=6$ to $N=21$, where $N=3L$.
The line fits the data to $a x^p+b$, where $a$, $b$ and $p$ are free parameters. The result gives an exponent $p=2.05\pm 0.05$.
The other parameters of the fit are $b=(1.0\pm0.5)\times 10^{-4}$ and $a=(158.7\pm 0.6)\times 10^{-3}$. Inset: the black points are $E_1-E_0$ as a function of $1/L$, 
as in the main panel. The red squares correspond to the difference between energy of the third excited and ground states, $E_3-E_0$. 
}\label{Fig2}
\end{figure}

We employ numerical methods to show that our previous result based on Majorana boundary modes is also valid for the full model Eqs.~(\ref{eq:3Isinga},\ref{eq:3Isingb}).
Specifically, exact diagonalizations of the three Ising coupled chains are carried out for sizes ranging from $N=6$ to $N=24$.
We use ARPACK libraries as only a few eigenstates are required~\cite{Le98}. These libraries are based on Lanczos algorithm which is specially suited to 
perform  efficient partial diagonalization of sparse matrices.

In Fig.\ \ref{Fig2}, the difference in energy between the first excited and ground-state, $E_1-E_0$, appears as a function of $1/L$.
The data have been fitted to a function $a * x^p+b$, where $a,b,p$ are free parameters. The result is the solid black line with
$p=2.05\pm 0.05$ and $b=(1.0\pm0.5)\times 10^{-4}$. These values are in excellent agreement with our 
previous analysis based on RKKY interactions between Majorana particles.

The low-energy doublet can be resolved in experiments only if higher excited states have much larger energy.
Of course, this occurs in the thermodynamic limit, as the doublet energy $E_{1}-E_0\sim 1/L^2$ is much smaller than the energy of the second excited 
state $E_{2}-E_0\sim 1/L$.
The red squares in the inset of Fig.\ \ref{Fig2} represent the difference in energy between the second-excited and ground-state, $E_2-E_0$, 
as a function of $1/L$. The black dots are  $E_1-E_0$ as in the main panel. We can see that $E_2-E_0$ is more than two orders of magnitude larger than $E_1-E_0$.
The size of the chains that we have employed runs from $L=2$ to $L=8$.
We conclude that Majorana modes are well protected even for small chains.

\begin{figure}[t]
\begin{center}
 \includegraphics[width=8.cm]{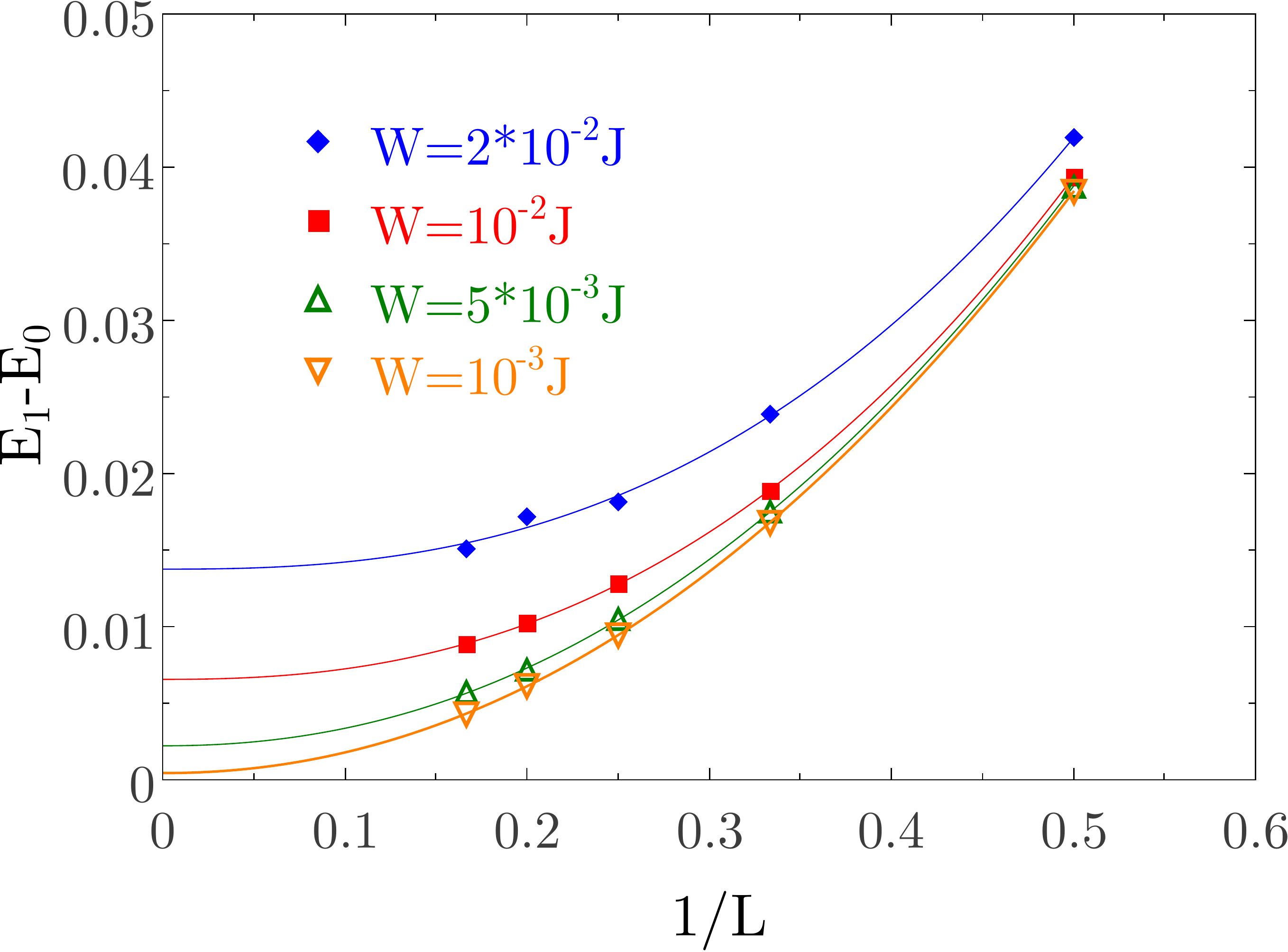}     
\end{center}
\caption{Difference between the energy of the first excited and ground states, $E_1-E_0$, as a function of the inverse of the chains length $1/L$.  
A random magnetic field in the $x$-axis, following a  uniform distribution in the interval $[-W/2,W/2]$, 
is added to the original Hamiltonian with $h=J=J_{c}$. 
Each set of data corresponds to a different value of $W$, as indicated in the legend.
The lines fit the points for each disorder to a function $a x^p+b$, where $a$, $b$ and $p$ are free parameters. 
The value of the fitting at the origin range approximately from $b=0.014$ to $b=0.0004$ and the exponent from $p=2.6$ to $p=2.07$.
}\label{Fig3}
\end{figure}

We discuss the source of incoherent dynamic in the experimental set-up. 
Fluctuations due to the quantum nature of the junctions can be kept small by employing a large ratio $E_J/E_C\gtrsim  100$\ \cite{BeSa12}.
In this case, phase slips are neither important as their probability decrease exponentially in $E_J/E_C$\ \cite{BePa14}.

The largest source of noise is caused by time-dependent fluctuations in the magnetic field threading the large loops $\varphi_1$, $\varphi_2$ (Fig.\ \ref{Fig1}a),
which produce additional contributions in the phases of ''small'' junctions of the order of $\sim \varphi_1/L$. 
These can be taken into accounts by introducing a small random argument in the third cosine of Eq.\ (\ref{eq:JE})
or, in the language of our effective spin model, by adding a small random magnetic field in the x-axis. 
It is important to realize that this noise can be minimized by constructing the JJA with a small area between nearby chains. 
We are going to show that Majorana modes are still protected if this noise is kept small enough.

We present results for exact diagonalization of our model in the presence of a random magnetic field in the x-axis.
In Fig.\ \ref{Fig3}, the energy difference of the first two levels $\Delta E_1$ is computed as a function of $1/L$ for the three Ising chains, 
Eq.~(\ref{eq:3Isinga},\ref{eq:3Isingb}), plus a small perturbation $\Delta H_x=\sum_{k} \epsilon_p(k) \sigma^{x}_p(k)$, 
where $\epsilon_p(k)$ are randomly distributed in the interval $[-W/2,W/2]$.
The solid lines fit each set of data to a function $a*x^p+b$, with free parameters $a,b,c$. For the largest disorder $W=2*10^{-2}J$, the fitting function
provides $p=2.6$ and it does not go trough the origin $b=0.014$. This is expected as the typical energy of the perturbation, $W$, is roughly the same 
as the one of the first level $\Delta E$ for the largest size. The points for the smaller disorder display a more similar behavior to the one in the clean system.
Furthermore, the low energy mode remains in all cases well separated from the bulk spectrum in the presence of small disorder.

\emph{Summary.} 
We have shown that Majorana modes appear in a system composed by three ladders of Josephson junctions.
The signature of Majorana is a low energy mode with energy going as $\sim 1/L^2$, while the bulk spectrum eigenstates have energies $\epsilon\sim 1/L$.
The lowest energy doublet is then protected in the thermodynamic limit.
We have further checked that the energy difference of the lowest doublet is more than two order of magnitude smaller than energy of bulk eigenstates
for chains of length $L\sim 10$. So, boundary modes are also protected for system sizes which are accessible to experiments.
Finally, we have analyzed the sources of incoherence and noise in the proposed experimental set-up. 
We conclude that our JJA can be a good experimental system to search for the Majorana particle.

M. P. and L. B. I. acknowledge support from ARO (W911NF-13-1-0431) and ANR QuDec. A. M. T. was supported by the U.S. Department of Energy (DOE), Division of Materials Science, under Contract No. DE-AC02-98CH10886.

\bibliographystyle{apsrev4-1}
\bibliography{./3Ising}

\begin{thebibliography}{19}%
\makeatletter
\providecommand \@ifxundefined [1]{%
 \@ifx{#1\undefined}
}%
\providecommand \@ifnum [1]{%
 \ifnum #1\expandafter \@firstoftwo
 \else \expandafter \@secondoftwo
 \fi
}%
\providecommand \@ifx [1]{%
 \ifx #1\expandafter \@firstoftwo
 \else \expandafter \@secondoftwo
 \fi
}%
\providecommand \natexlab [1]{#1}%
\providecommand \enquote  [1]{``#1''}%
\providecommand \bibnamefont  [1]{#1}%
\providecommand \bibfnamefont [1]{#1}%
\providecommand \citenamefont [1]{#1}%
\providecommand \href@noop [0]{\@secondoftwo}%
\providecommand \href [0]{\begingroup \@sanitize@url \@href}%
\providecommand \@href[1]{\@@startlink{#1}\@@href}%
\providecommand \@@href[1]{\endgroup#1\@@endlink}%
\providecommand \@sanitize@url [0]{\catcode `\\12\catcode `\$12\catcode
  `\&12\catcode `\#12\catcode `\^12\catcode `\_12\catcode `\%12\relax}%
\providecommand \@@startlink[1]{}%
\providecommand \@@endlink[0]{}%
\providecommand \url  [0]{\begingroup\@sanitize@url \@url }%
\providecommand \@url [1]{\endgroup\@href {#1}{\urlprefix }}%
\providecommand \urlprefix  [0]{URL }%
\providecommand \Eprint [0]{\href }%
\providecommand \doibase [0]{http://dx.doi.org/}%
\providecommand \selectlanguage [0]{\@gobble}%
\providecommand \bibinfo  [0]{\@secondoftwo}%
\providecommand \bibfield  [0]{\@secondoftwo}%
\providecommand \translation [1]{[#1]}%
\providecommand \BibitemOpen [0]{}%
\providecommand \bibitemStop [0]{}%
\providecommand \bibitemNoStop [0]{.\EOS\space}%
\providecommand \EOS [0]{\spacefactor3000\relax}%
\providecommand \BibitemShut  [1]{\csname bibitem#1\endcsname}%
\let\auto@bib@innerbib\@empty
\bibitem [{\citenamefont {Majorana}(1937)}]{Ma37}%
  \BibitemOpen
  \bibfield  {author} {\bibinfo {author} {\bibfnamefont {E.}~\bibnamefont
  {Majorana}},\ }\href {\doibase 10.1007/BF02961314} {\bibfield  {journal}
  {\bibinfo  {journal} {Il Nuovo Cimento}\ }\textbf {\bibinfo {volume} {14}},\
  \bibinfo {pages} {171} (\bibinfo {year} {1937})}\BibitemShut {NoStop}%
\bibitem [{\citenamefont {Alicea}(2012)}]{Al12}%
  \BibitemOpen
  \bibfield  {author} {\bibinfo {author} {\bibfnamefont {J.}~\bibnamefont
  {Alicea}},\ }\href@noop {} {\bibfield  {journal} {\bibinfo  {journal}
  {Reports on Progress in Physics}\ }\textbf {\bibinfo {volume} {75}},\
  \bibinfo {pages} {076501} (\bibinfo {year} {2012})}\BibitemShut {NoStop}%
\bibitem [{\citenamefont {Kitaev}(2001)}]{Ki01}%
  \BibitemOpen
  \bibfield  {author} {\bibinfo {author} {\bibfnamefont {A.~Y.}\ \bibnamefont
  {Kitaev}},\ }\href {http://stacks.iop.org/1063-7869/44/i=10S/a=S29}
  {\bibfield  {journal} {\bibinfo  {journal} {Physics-Uspekhi}\ }\textbf
  {\bibinfo {volume} {44}},\ \bibinfo {pages} {131} (\bibinfo {year}
  {2001})}\BibitemShut {NoStop}%
\bibitem [{\citenamefont {Mourik}\ \emph {et~al.}(2012)\citenamefont {Mourik},
  \citenamefont {Zuo}, \citenamefont {Frolov}, \citenamefont {Plissard},
  \citenamefont {Bakkers},\ and\ \citenamefont {Kouwenhoven}}]{Mo12}%
  \BibitemOpen
  \bibfield  {author} {\bibinfo {author} {\bibfnamefont {V.}~\bibnamefont
  {Mourik}}, \bibinfo {author} {\bibfnamefont {K.}~\bibnamefont {Zuo}},
  \bibinfo {author} {\bibfnamefont {S.}~\bibnamefont {Frolov}}, \bibinfo
  {author} {\bibfnamefont {S.}~\bibnamefont {Plissard}}, \bibinfo {author}
  {\bibfnamefont {E.}~\bibnamefont {Bakkers}}, \ and\ \bibinfo {author}
  {\bibfnamefont {L.}~\bibnamefont {Kouwenhoven}},\ }\href@noop {} {\bibfield
  {journal} {\bibinfo  {journal} {Science}\ }\textbf {\bibinfo {volume}
  {336}},\ \bibinfo {pages} {1003} (\bibinfo {year} {2012})}\BibitemShut
  {NoStop}%
\bibitem [{\citenamefont {Hasan}\ and\ \citenamefont {Kane}(2010)}]{Ha10}%
  \BibitemOpen
  \bibfield  {author} {\bibinfo {author} {\bibfnamefont {M.~Z.}\ \bibnamefont
  {Hasan}}\ and\ \bibinfo {author} {\bibfnamefont {C.~L.}\ \bibnamefont
  {Kane}},\ }\href {\doibase 10.1103/RevModPhys.82.3045} {\bibfield  {journal}
  {\bibinfo  {journal} {Rev. Mod. Phys.}\ }\textbf {\bibinfo {volume} {82}},\
  \bibinfo {pages} {3045} (\bibinfo {year} {2010})}\BibitemShut {NoStop}%
\bibitem [{\citenamefont {Bell}\ \emph
  {et~al.}(2012{\natexlab{a}})\citenamefont {Bell}, \citenamefont {Ioffe},\
  and\ \citenamefont {Gershenson}}]{BeIo12}%
  \BibitemOpen
  \bibfield  {author} {\bibinfo {author} {\bibfnamefont {M.~T.}\ \bibnamefont
  {Bell}}, \bibinfo {author} {\bibfnamefont {L.~B.}\ \bibnamefont {Ioffe}}, \
  and\ \bibinfo {author} {\bibfnamefont {M.~E.}\ \bibnamefont {Gershenson}},\
  }\href {\doibase 10.1103/PhysRevB.86.144512} {\bibfield  {journal} {\bibinfo
  {journal} {Phys. Rev. B}\ }\textbf {\bibinfo {volume} {86}},\ \bibinfo
  {pages} {144512} (\bibinfo {year} {2012}{\natexlab{a}})}\BibitemShut
  {NoStop}%
\bibitem [{\citenamefont {Bell}\ \emph {et~al.}(2014)\citenamefont {Bell},
  \citenamefont {Paramanandam}, \citenamefont {Ioffe},\ and\ \citenamefont
  {Gershenson}}]{BePa14}%
  \BibitemOpen
  \bibfield  {author} {\bibinfo {author} {\bibfnamefont {M.~T.}\ \bibnamefont
  {Bell}}, \bibinfo {author} {\bibfnamefont {J.}~\bibnamefont {Paramanandam}},
  \bibinfo {author} {\bibfnamefont {L.~B.}\ \bibnamefont {Ioffe}}, \ and\
  \bibinfo {author} {\bibfnamefont {M.~E.}\ \bibnamefont {Gershenson}},\ }\href
  {\doibase 10.1103/PhysRevLett.112.167001} {\bibfield  {journal} {\bibinfo
  {journal} {Phys. Rev. Lett.}\ }\textbf {\bibinfo {volume} {112}},\ \bibinfo
  {pages} {167001} (\bibinfo {year} {2014})}\BibitemShut {NoStop}%
\bibitem [{\citenamefont {Bell}\ \emph
  {et~al.}(2012{\natexlab{b}})\citenamefont {Bell}, \citenamefont {Sadovskyy},
  \citenamefont {Ioffe}, \citenamefont {Kitaev},\ and\ \citenamefont
  {Gershenson}}]{BeSa12}%
  \BibitemOpen
  \bibfield  {author} {\bibinfo {author} {\bibfnamefont {M.~T.}\ \bibnamefont
  {Bell}}, \bibinfo {author} {\bibfnamefont {I.~A.}\ \bibnamefont {Sadovskyy}},
  \bibinfo {author} {\bibfnamefont {L.~B.}\ \bibnamefont {Ioffe}}, \bibinfo
  {author} {\bibfnamefont {A.~Y.}\ \bibnamefont {Kitaev}}, \ and\ \bibinfo
  {author} {\bibfnamefont {M.~E.}\ \bibnamefont {Gershenson}},\ }\href@noop {}
  {\bibfield  {journal} {\bibinfo  {journal} {Phys. Rev. Lett.}\ }\textbf
  {\bibinfo {volume} {109}},\ \bibinfo {pages} {137003} (\bibinfo {year}
  {2012}{\natexlab{b}})}\BibitemShut {NoStop}%
\bibitem [{\citenamefont {Crampé}\ and\ \citenamefont
  {Trombettoni}(2013)}]{Cr13}%
  \BibitemOpen
  \bibfield  {author} {\bibinfo {author} {\bibfnamefont {N.}~\bibnamefont
  {Crampé}}\ and\ \bibinfo {author} {\bibfnamefont {A.}~\bibnamefont
  {Trombettoni}},\ }\href {\doibase
  http://dx.doi.org/10.1016/j.nuclphysb.2013.03.001} {\bibfield  {journal}
  {\bibinfo  {journal} {Nuclear Physics B}\ }\textbf {\bibinfo {volume}
  {871}},\ \bibinfo {pages} {526 } (\bibinfo {year} {2013})}\BibitemShut
  {NoStop}%
\bibitem [{\citenamefont {Tsvelik}(2013)}]{Ts13}%
  \BibitemOpen
  \bibfield  {author} {\bibinfo {author} {\bibfnamefont {A.~M.}\ \bibnamefont
  {Tsvelik}},\ }\href {\doibase 10.1103/PhysRevLett.110.147202} {\bibfield
  {journal} {\bibinfo  {journal} {Phys. Rev. Lett.}\ }\textbf {\bibinfo
  {volume} {110}},\ \bibinfo {pages} {147202} (\bibinfo {year}
  {2013})}\BibitemShut {NoStop}%
\bibitem [{\citenamefont {Coleman}\ \emph {et~al.}(1995)\citenamefont
  {Coleman}, \citenamefont {Ioffe},\ and\ \citenamefont {Tsvelik}}]{Co95}%
  \BibitemOpen
  \bibfield  {author} {\bibinfo {author} {\bibfnamefont {P.}~\bibnamefont
  {Coleman}}, \bibinfo {author} {\bibfnamefont {L.~B.}\ \bibnamefont {Ioffe}},
  \ and\ \bibinfo {author} {\bibfnamefont {A.~M.}\ \bibnamefont {Tsvelik}},\
  }\href {\doibase 10.1103/PhysRevB.52.6611} {\bibfield  {journal} {\bibinfo
  {journal} {Phys. Rev. B}\ }\textbf {\bibinfo {volume} {52}},\ \bibinfo
  {pages} {6611} (\bibinfo {year} {1995})}\BibitemShut {NoStop}%
\bibitem [{\citenamefont {Altland}\ \emph {et~al.}(2014)\citenamefont
  {Altland}, \citenamefont {B\'eri}, \citenamefont {Egger},\ and\ \citenamefont
  {Tsvelik}}]{Al14}%
  \BibitemOpen
  \bibfield  {author} {\bibinfo {author} {\bibfnamefont {A.}~\bibnamefont
  {Altland}}, \bibinfo {author} {\bibfnamefont {B.}~\bibnamefont {B\'eri}},
  \bibinfo {author} {\bibfnamefont {R.}~\bibnamefont {Egger}}, \ and\ \bibinfo
  {author} {\bibfnamefont {A.}~\bibnamefont {Tsvelik}},\ }\href {\doibase
  10.1103/PhysRevLett.113.076401} {\bibfield  {journal} {\bibinfo  {journal}
  {Phys. Rev. Lett.}\ }\textbf {\bibinfo {volume} {113}},\ \bibinfo {pages}
  {076401} (\bibinfo {year} {2014})}\BibitemShut {NoStop}%
\bibitem [{\citenamefont {Giuliano}\ and\ \citenamefont {Sodano}(2013)}]{Gi13}%
  \BibitemOpen
  \bibfield  {author} {\bibinfo {author} {\bibfnamefont {D.}~\bibnamefont
  {Giuliano}}\ and\ \bibinfo {author} {\bibfnamefont {P.}~\bibnamefont
  {Sodano}},\ }\href {http://stacks.iop.org/0295-5075/103/i=5/a=57006}
  {\bibfield  {journal} {\bibinfo  {journal} {EPL (Europhysics Letters)}\
  }\textbf {\bibinfo {volume} {103}},\ \bibinfo {pages} {57006} (\bibinfo
  {year} {2013})}\BibitemShut {NoStop}%
\bibitem [{\citenamefont {Ruderman}\ and\ \citenamefont {Kittel}(1954)}]{Ru54}%
  \BibitemOpen
  \bibfield  {author} {\bibinfo {author} {\bibfnamefont {M.~A.}\ \bibnamefont
  {Ruderman}}\ and\ \bibinfo {author} {\bibfnamefont {C.}~\bibnamefont
  {Kittel}},\ }\href {\doibase 10.1103/PhysRev.96.99} {\bibfield  {journal}
  {\bibinfo  {journal} {Phys. Rev.}\ }\textbf {\bibinfo {volume} {96}},\
  \bibinfo {pages} {99} (\bibinfo {year} {1954})}\BibitemShut {NoStop}%
\bibitem [{\citenamefont {Kasuya}(1956)}]{Ka56}%
  \BibitemOpen
  \bibfield  {author} {\bibinfo {author} {\bibfnamefont {T.}~\bibnamefont
  {Kasuya}},\ }\href {\doibase 10.1143/PTP.16.45} {\bibfield  {journal}
  {\bibinfo  {journal} {Progress of Theoretical Physics}\ }\textbf {\bibinfo
  {volume} {16}},\ \bibinfo {pages} {45} (\bibinfo {year} {1956})}\BibitemShut
  {NoStop}%
\bibitem [{\citenamefont {Yosida}(1957)}]{Yo57}%
  \BibitemOpen
  \bibfield  {author} {\bibinfo {author} {\bibfnamefont {K.}~\bibnamefont
  {Yosida}},\ }\href {\doibase 10.1103/PhysRev.106.893} {\bibfield  {journal}
  {\bibinfo  {journal} {Phys. Rev.}\ }\textbf {\bibinfo {volume} {106}},\
  \bibinfo {pages} {893} (\bibinfo {year} {1957})}\BibitemShut {NoStop}%
\bibitem [{\citenamefont {Eriksson}\ \emph {et~al.}(2015)\citenamefont
  {Eriksson}, \citenamefont {Zazunov}, \citenamefont {Sodano},\ and\
  \citenamefont {Egger}}]{Er15}%
  \BibitemOpen
  \bibfield  {author} {\bibinfo {author} {\bibfnamefont {E.}~\bibnamefont
  {Eriksson}}, \bibinfo {author} {\bibfnamefont {A.}~\bibnamefont {Zazunov}},
  \bibinfo {author} {\bibfnamefont {P.}~\bibnamefont {Sodano}}, \ and\ \bibinfo
  {author} {\bibfnamefont {R.}~\bibnamefont {Egger}},\ }\href {\doibase
  10.1103/PhysRevB.91.064501} {\bibfield  {journal} {\bibinfo  {journal} {Phys.
  Rev. B}\ }\textbf {\bibinfo {volume} {91}},\ \bibinfo {pages} {064501}
  (\bibinfo {year} {2015})}\BibitemShut {NoStop}%
\bibitem [{\citenamefont {Pfeuty}(1970)}]{Pf70}%
  \BibitemOpen
  \bibfield  {author} {\bibinfo {author} {\bibfnamefont {P.}~\bibnamefont
  {Pfeuty}},\ }\href {\doibase http://dx.doi.org/10.1016/0003-4916(70)90270-8}
  {\bibfield  {journal} {\bibinfo  {journal} {Annals of Physics}\ }\textbf
  {\bibinfo {volume} {57}},\ \bibinfo {pages} {79 } (\bibinfo {year}
  {1970})}\BibitemShut {NoStop}%
\bibitem [{\citenamefont {Lehoucq}\ \emph {et~al.}(1998)\citenamefont
  {Lehoucq}, \citenamefont {Sorensen},\ and\ \citenamefont {Yang}}]{Le98}%
  \BibitemOpen
  \bibfield  {author} {\bibinfo {author} {\bibfnamefont {R.~B.}\ \bibnamefont
  {Lehoucq}}, \bibinfo {author} {\bibfnamefont {D.~C.}\ \bibnamefont
  {Sorensen}}, \ and\ \bibinfo {author} {\bibfnamefont {C.}~\bibnamefont
  {Yang}},\ }\href@noop {} {\emph {\bibinfo {title} {ARPACK users' guide:
  solution of large-scale eigenvalue problems with implicitly restarted Arnoldi
  methods}}},\ Vol.~\bibinfo {volume} {6}\ (\bibinfo  {publisher} {Siam},\
  \bibinfo {year} {1998})\BibitemShut {NoStop}%
\end{thebibliography}%
\end{document}